# EVALUATION OF THE MOMENTS OF INERTIA OF FORCED SPLIT FRAGMENTS FOR $^{232}\mathrm{Th}(n,f)$ AND $^{238}\mathrm{U}(n,f)$ NUCLEI

D. E. Lyubashevsky[1], P. V. Kostryukov[1,2], J. D. Shcherbina[1],

T. Yu. Shashkina[1], S. V. Klyuchnikov[1]

1Voronezh State University Faculty of Physics Universitetskaya pl. 1, Voronezh Voronezh Oblast 394006 Russian Federation

2Voronezh State University of Forestry and Technologies named after G F Morozov ul. Timiryazeva, 8, Voronezh 394087 Russian Federation

This study develops an innovative theoretical framework that integrates macroscopic liquid-drop model with microscopic superfluid theory to calculate moments of inertia for fission fragments, extending our previous spontaneous fission approach to include neutron-induced threshold fission of $^{232}\mathrm{Th}(n,f)$ and $^{238}\mathrm{U}(n,f)$. The model provides a comprehensive description of fission dynamics by simultaneously accounting for collective vibrational modes (bending and wriggling) and their influence on spin distributions, while systematically investigating the deformation dependence of moments of inertia. Our calculations demonstrate good agreement with experimental data, validating the model's reliability for both fundamental nuclear fission studies and practical applications in reactor physics. The unified treatment of macroscopic and microscopic effects offers new insights into fission mechanisms and enables accurate predictions of fragment characteristics across the entire mass range. These results provide a solid basis for future studies of exotic fission processes and advanced applications in nuclear energy. The methodological advances presented here open new possibilities for theoretical studies of various heavy-ion reactions and fission phenomena in superheavy nuclei.

## INTRODUCTION

Modern nuclear fission studies combine experimental [1] and theoretical [2,3] approaches to investigate binary fission of $^{232}\mathrm{Th}(n,f)$ and $^{238}\mathrm{U}(n,f)$ isotopes induced by threshold-energy neutrons. This process involves the formation of a compound nucleus undergoing fission that decays into two significantly distinct fragments – light and heavy – characterized by different charges, masses, spins, and other physical parameters.
The analysis of these disparities is crucial for understanding fission mechanisms as it enables establishing correlations between the parent
nucleus structure, excitation energy, deformation dynamics, and final fragment cha racteristics.

A particularly important aspect of contemporary research involves studying quantum mechanical interference effects, which are critical for describing fundamental binary fission characteristics. An adequate theoretical description of these phenomena requires quantum approaches based on analyzing nuclear and particle wave functions at different stages of the fission process. Significant contributions to this theory have been made by both international [4] and domestic [5] researchers, whose work has shaped modern understanding of quantum aspects in nuclear processes.

Nuclear fission represents a multistage dynamic process in which various degrees of freedom are sequentially activated, determining key fragment characteristics: spin distributions (SDs), energy spectra, and kinematic parameters. The central phenomenon in this process involves nonequilibrium deformations of the nucleus arising from substantial nonlinear deviations from equilibrium configurations. These deformations lead to accumulation of significant excitation energy (up to tens of MeV), subsequently redistributed within the system and determining such important characteristics as prompt neutron multiplicity and features of their spin state distributions.

The terminal stage of nuclear fission is characterized by intricate energy redistribution processes and stabilization of resulting fragments. During this phase, the excess energy is released mainly through two channels: neutron evaporation and γ-quantum emission. These processes, governed by fundamental nuclear physics principles, reflect specific features of nucleon-nucleon interactions and angular momentum redistribution dynamics within the system. The cascade particle evaporation serves as an efficient mechanism for dissipating excitation energy, progressively transitioning fragments to minimal-energy states – either ground or weakly excited states.

The pivotal stage of the entire process culminates with the completion of cascade transitions, as the fragment system attains thermodynamic equilibrium. At this juncture, the stable configurations of the fission fragments become definitively established, with their spin characteristics—both magnitude and spatial orientation—becoming fixed and remaining invariant thereafter. This phase holds fundamental significance, as it precisely determines the ultimate spin states of the fragments, which subsequently govern all their physical properties and interaction modalities with surrounding nuclear and atomic systems. The complete stabilization of spin parameters signifies the conclusion of the dynamic fission phase and the system's transition to an equilibrium state.

These investigations significantly improve our understanding of the fundamental principles of nuclear fission, spanning from the initial scission of the nuclear system to the formation of final reaction products. The obtained results carry paramount importance for advancing contemporary theories of nuclear reactions and developing more accurate predictive models, finding applications in both fundamental science and applied nuclear technologies. Of particular value is the capacity to correlate theoretical predictions with experimental data on spin characteristics, serving as a crucial validation criterion for the developed theoretical frameworks.

This study extends the authors' previously developed spontaneous fission model to describe neutron-induced fission of $^{232}\mathrm{Th}(n,f)$ and $^{238}\mathrm{U}(n,f)$ isotopes at threshold energies. Primary emphasis is placed on detailed analysis of non-equilibrium fragment deformations and computation of their moments of inertia employing modern high-precision theoretical methodologies. Particular importance is accorded to model verification through meticulous comparison of theoretical predictions with experimental data, which will not only validate the adequacy of the developed approach but also yield novel fundamental insights into fission dynamics.

An essential component of this work involves investigating the model's predictive capability regarding fragment SDs. Such analysis unveils prospects for enhanced understanding of nuclear fission mechanisms, including secondary product formation processes and specific features of energy redistribution within the system. The obtained results possess substantial potential for advancing nuclear reaction theory and may find applications across various domains of nuclear physics.

## METHODS OF ESTIMATION OF MOMENTS OF INERTIA

### A. A model of a "cold" fission system

This study investigates the hypothesis of a sawtooth dependence of the moments of inertia on the mass number, representing a key avenue for further elucidation of the internal mechanisms governing nuclear structure and fission behaviour. Understanding these dependencies is essential for a comprehensive understanding of the physical phenomena involved in the fission process, with significant implications for theoretical nuclear physics and practical applications in fields such as nuclear energy and medicine.

The process of binary threshold fission of a compound fission system (CFS) can be described using the quantum theory of fission based on the generalised model of the nucleus proposed in [4]. This model provides a versatile and robust framework that accounts for both nucleonic and collective degrees of freedom. These interactions, associated with nuclear deformation and vibrational dynamics, enable a deeper understanding and more accurate modelling of the dynamics of the fission process. Incorporating such factors can significantly enhances the precision of predictions, particularly for the complex and multifaceted mechanisms occurring in the nucleus during the fission. This approach facilitates comprehensive modeling of critical aspects, from the nuclear deformations to vibrational effects that influence the final outcome of the process.

Forced fission occurs when a target nucleus $(A,Z)$ captures a neutron with a kinetic energy $T_n$, corresponding to the threshold energy for a given parent nucleus. This process leads to excitation of the nucleus, which accumulates an excitation energy $B_n + T_n$, including both the contribution from the binding energy of the captured neutron $B_n$ (approximately 6 MeV) and the additional kinetic energy of the neutron $T_n$ (approximately 2 MeV). Within a timescale of $T_0 \approx 10^{-22}$ s the excited nucleus rapidly transitions to the neutron resonance CFS state. To describe the wave function $\psi_K^{JM}$ of this state, an approach based on Wigner's random matrix theory [6,7,8] is used, providing a precise framework for describing the quantum states of CFS:

$$\psi_K^{JM} = \sum_{i \neq 0} b_i \psi_{iK}^{JM} + b_0 \psi_{0K}^{JM}(\beta_\lambda), \quad (1)$$

in this theoretical model, the functions $\psi_{iK}^{JM}$ and $\psi_{0K}^{JM}(\beta_\lambda)$ represent components describing distinct aspects of the CFS dynamics. Specifically, $\psi_{iK}^{JM}$ is associated with $i$ -quasiparticle excited state of the system, and $\psi_{0K}^{JM}(\beta_\lambda)$ reflects the collective deformation motion of the CFS, which plays a pivotal role in the transient dividing state first introduced by A. Bohr [4]. The excitation energy $B_n + T_n$, associated with the state $\psi_{0K}^{JM}(\beta_\lambda)$ corresponds to the transient state of CFS, where the system takes the form corresponding to fission. The formula for the wave function in this model takes into account the contribution of these states through the squares of the coefficients $b_i$ and $b_0$, which are weighted averages $1/N$, where $N$ is the total number of quasiparticle states contributing to the wave function (1), for all quasiparticle states involved in the formation of the wave function.

The forced threshold fission of a CFS occurs with significant probability if the excitation energy $B_n + T_n$ exceeds the heights of the internal $B_1$ and external $B_2$ fission deformation barriers. This process is characteristic of target nuclei $^{238}U$ and $^{232}Th$, interacting with neutrons of threshold energies. This is supported by Figure 1, where the upper arrow indicates the critical energy at which the fission probability significantly increases.

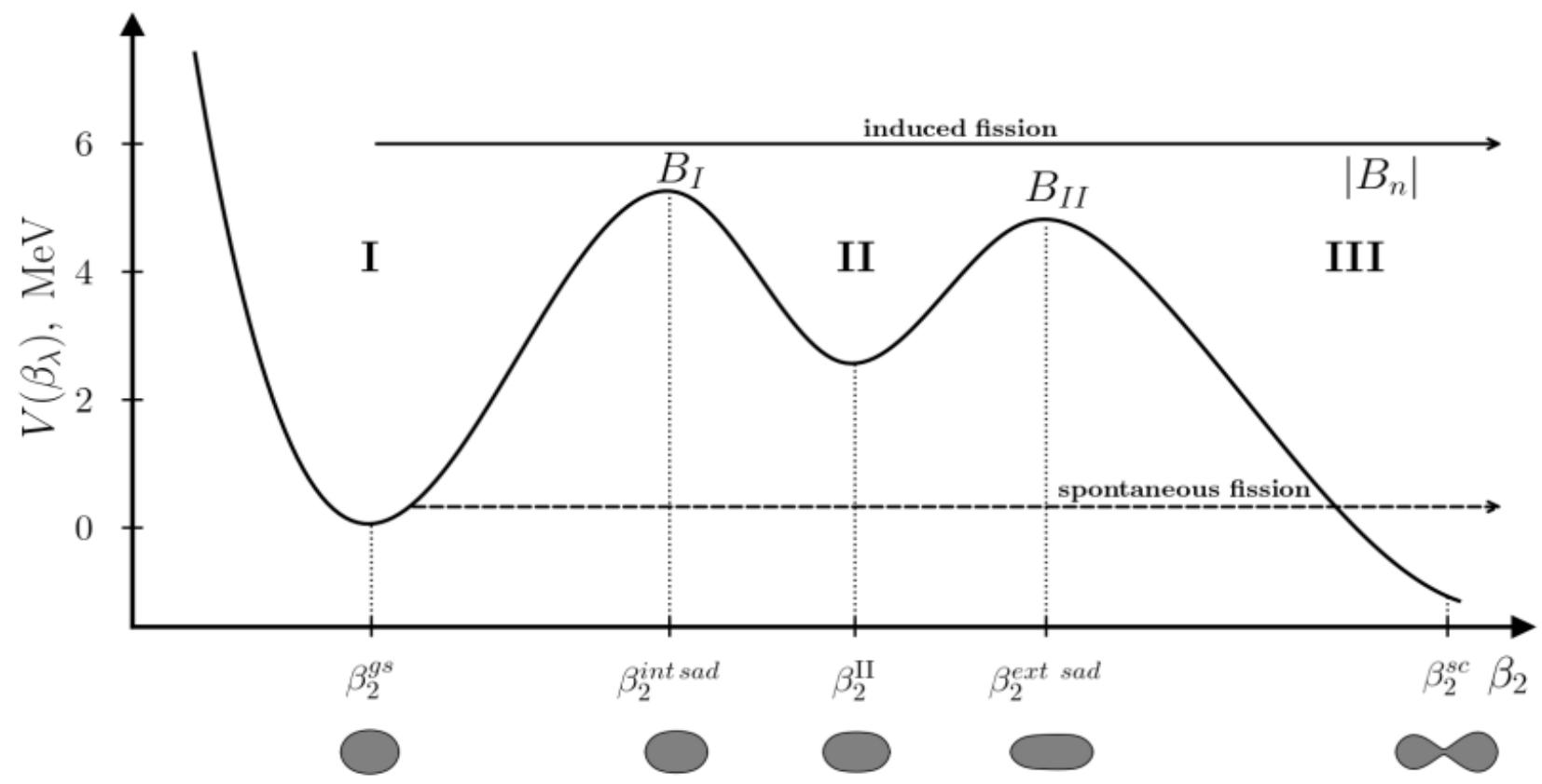

*Fig. 1. Principal diagram of the potential $V$ as a function of the quadrupole deformation of the nucleus $\beta_2$. Region I corresponds to the ground state of the nucleus with $\beta_2^{gs}$, II to the isomeric states, and III to the out-of-barrier region where the nucleus decays into fission fragments. $B_I$ refers to the internal barrier, $B_{II}$ refers to the external barrier.*

The parameters of forced threshold fission of CFS are described through two fundamental postulates outlined in the theoretical model [4]. The first postulate states asserts that the axial symmetry of the CFS is preserved during the fission process. This assumption is supported by empirical data demonstrating a high degree of fission symmetry in numerous of experimental studies. The second postulate posits that the projection $K$ of spin $J$ of the fissioning nucleus onto the symmetry axis remains constant throughout the process, beginning from the moment when the nucleus overcomes the external saddle state of the deformation potential. This hypothesis is crucial in theoretical analyses, since it allows to reliably describe the dynamics of fission and the spin characteristics of the system at various stages [9].

A key factor preventing the preservation of the spin projection on the symmetry axis during fission is the intense thermal excitation of both the fissile nucleus itself and the fragments formed. An increase in temperature enhances the dynamical Coriolis interaction, which, under conditions of thermal excitation, becomes dominant and significantly affects the system's behavior. According to theoretical studies [4, 10, 11], this effect leads to a statistical redistribution of all possible values of the projection $K$ of spin $J$ on the symmetry axis, especially at intermediate temperatures. This redistribution substantially influences the fission dynamics, reducing the distinctions between spin orientations and complicating accurate modeling of the system's behavior at different stages.

The statistical mixing of $K$ spin projections $J$ near the moment of fission of a CFS into fragments can be considered as a mechanism leading to a «loss of memory» about the initial values of these projections associated with transient fission states. This effect explains the absence of all types of asymmetries in the angular distributions of both binary and ternary nuclear fission products, including asymmetries related to odd and even values of spin projections [12] A similar phenomenon is observed for asymmetries with different parity with respect to the $P$ - и $T$-symmetry operators characterizing the properties of fission products, which is confirmed by experimental data [13, 14].

These experimental results indicate that the «cold» character of the system is preserved throughout all stages of the fission process. This holds true both for the stage of the downward motion of the nucleus from the outer saddle point of the deformation potential and for the formation of the angular distributions of fission products. Thus, the stability of the quasi-static characteristics of the system, even under conditions of low-temperature dynamics, confirms the critical role of slow processes of angular momentum redistribution in fission.

### B. Determination of nonequilibrium pre-fragment deformations

According to the postulate on the "cold" nature of a CFS discussed in the previous section, all excitation energy accumulated by pre-fragments during their formation is preferentially converted into collective deformation states. These states induce significant nonequilibrium deformations in the pre-fragments, which play a pivotal role in the subsequent fission dynamics. After the CFS rupture, the fission fragments undergo termalization, followed by a de-excitation process accompanied by neutron emission over a characteristic timescale $\tau_{nuc}$. This sequential transition between stages enables a quantitative assessment of the fragments' excitation levels. Of particular significance in this context is the work of V. Strutinsky [15], which established the relationship between the collective deformations of fragments and their excitation energy within the framework of the liquid-drop model of the nucleus. It is shown that this dependence allows us not only to characterize pre-fragments by their excited state, but also to determine a quantitative relationship between the degree of nonequilibrium deformation and the excitation energy, as well as the number of neutrons emitted by fragments during their thermal relaxation. These results not only deepen our understanding of the physical processes occurring during the fission stages, but also lay the foundation for refining models of the energy and angular distribution of fission products.

Data on the excitation energies of nuclei $^{238}\mathrm{U}\,(nf)$ and $^{232}\mathrm{Th}\,(nf)$, formed immediately after the fission system ruptures, are essential. An extensive analysis of literature sources provided the necessary data on excitation energies for the reaction of forced fission induced by neutrons at threshold energies, as presented in [16]. In particular, Figure 2 of this publication shows the excitation energies of fission fragments as a function of their mass number $A$, which allows us to use these data for the subsequent analysis of nonequilibrium deformations.

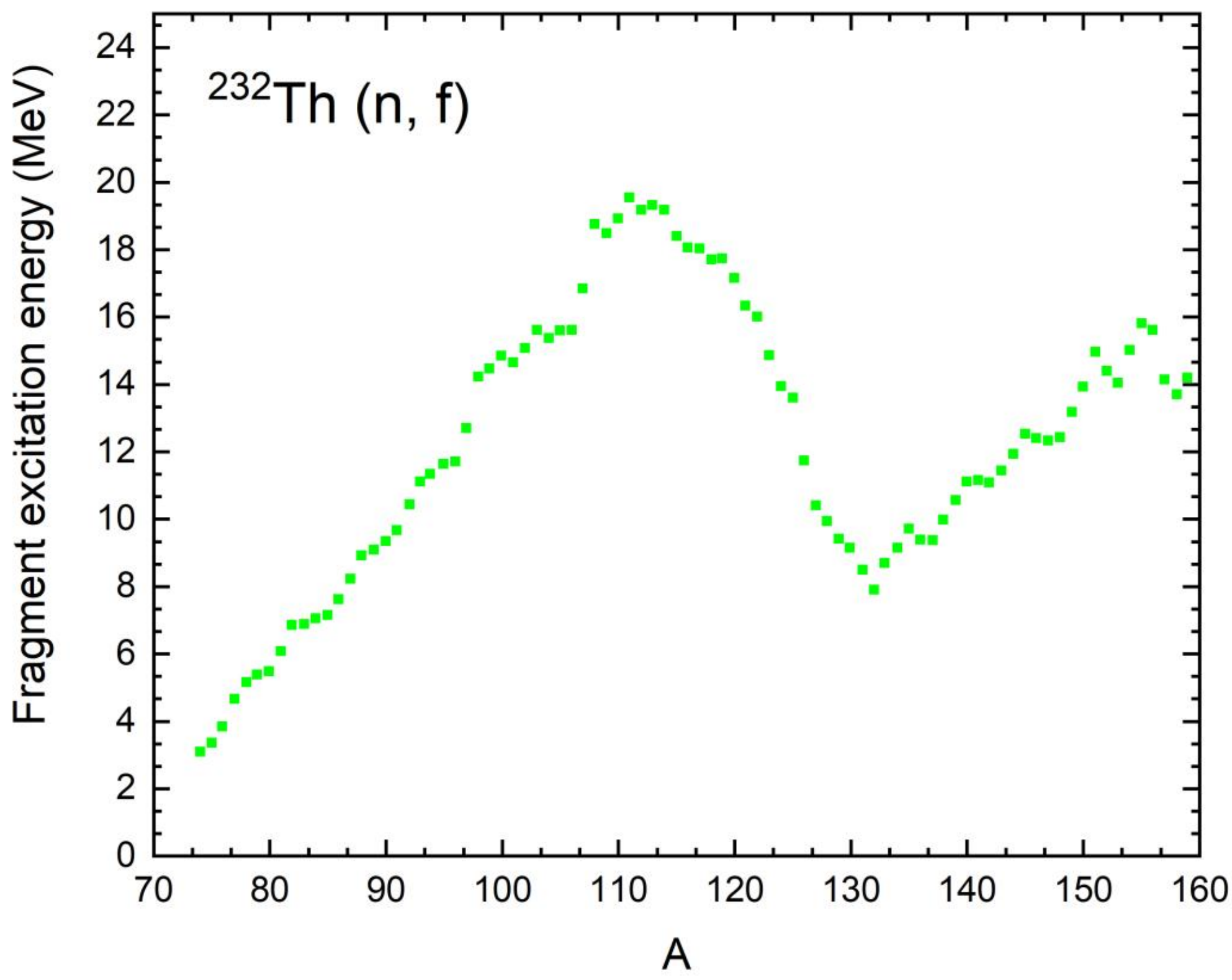

*Fig. 2 Dependence of the average excitation energies on the mass of fragments produced during forced threshold fission of* $^{232}\mathrm{Th}\,(n,f)$ *by neutrons with a kinetic energy of 2 MeV [16] .*

However, despite an extensive search in the available literature, it was not possible to find data on excitation energies for fragments produced in the forced fission of $^{238}\mathrm{U}$ nuclei by neutrons at threshold energies were not found, and only neutron yields for this nucleus were reported in [17]. Consequently, a methodologically sound approach is required to analyze neutron yields and their relationship to excitation energies. This study focuses on applying the approaches described in [18] to analyze neutron yields, conduct an in-depth analysis, and achieve a more accurate interpretation of the results. Two theoretical approaches from [18] are considered: the first employs the FREYA software package [19], and the second is based on a theoretical analysis of fission fragment decay, incorporating strict conservation of total angular momentum and parity, as proposed in [20]. Both approaches, when applied to spontaneous fission of $^{252}\mathrm{Cf}$, showed only qualitative agreement but lacked reasonable quantitative agreement, prompting the use experimental data from [21]. In the work, we directly utilize neutron yields from [17] presented in Fig. 3 for the case of forced fission induced by neutrons at threshold energies for the nucleus $^{238}\mathrm{U}\left(n,f\right)$.

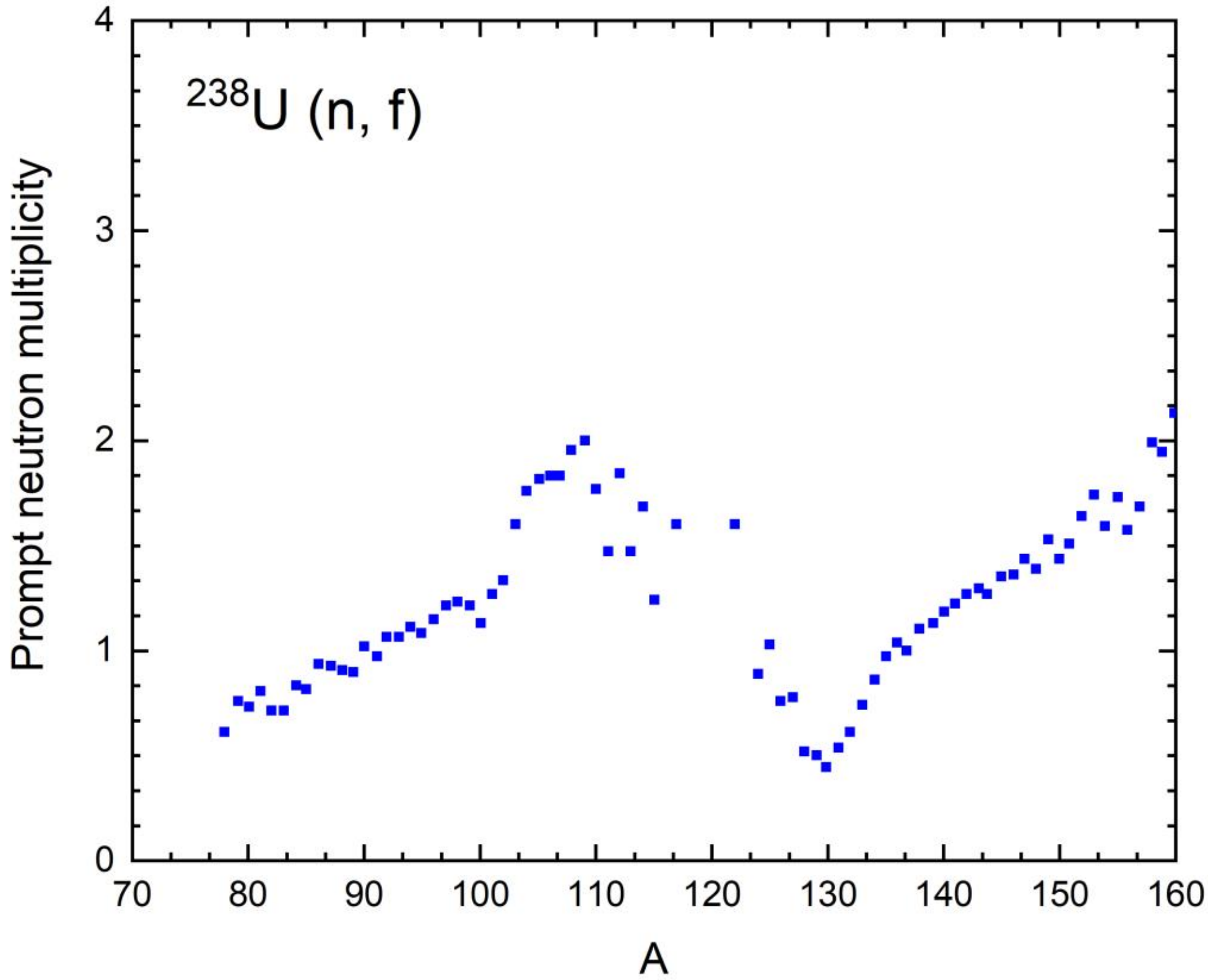

*Fig. 3. Dependence of the average neutron multiplicity on the mass of fragments produced during forced threshold fission of* $^{238}\mathrm{U}\left(n,f\right)$ *by neutrons with a kinetic energy of 2 MeV [17].*

Having determined the neutron yields, we proceed to calculate the excitation energy $U$ there are a number of approaches, each based on different theoretical assumptions and empirical data. One such method was proposed about a quarter of a century ago [20], provides a conceptual framework that establishes a quantitative relationship between excitation energy $U$ and neutron yield multiplicity. Within this methodology, this relationship is expressed as:

$$U = 5 + 4\nu + \nu^2, \quad (2)$$

where $\nu$ represents the neutron multiplicity.

Another advanced method, based on theoretical developments presented in [22], offers an improved model for calculating $U$. This approach extends the existing conceptual framework by incorporating more complex interaction mechanisms and relationships between key physical parameters that determine the dynamics of the fission process. The methodology includes a detailed consideration of collective and single-quasiparticle degrees of freedom, as well as the influence of different deformation configurations on the excitation energy. Consequently, a mathematical dependence is formalised, expressed by

$$U = 7(\nu + 3/7), \quad (3)$$

The linear dependence indicates that as the neutron multiplicity increases, the average excitation energy also increases. This phenomenon arises because the residual nucleus approaches a state near the stability line.

Using equation (2), we calculated the average excitation energy $U$ for the nucleus $^{238}\mathrm{U}(n,f)$, as shown in Figure 4, green solid line. In parallel, applying equation (3), the averaged value of the excitation energy $U$, which are represented in Figure 4 by the yellow line is calculated.

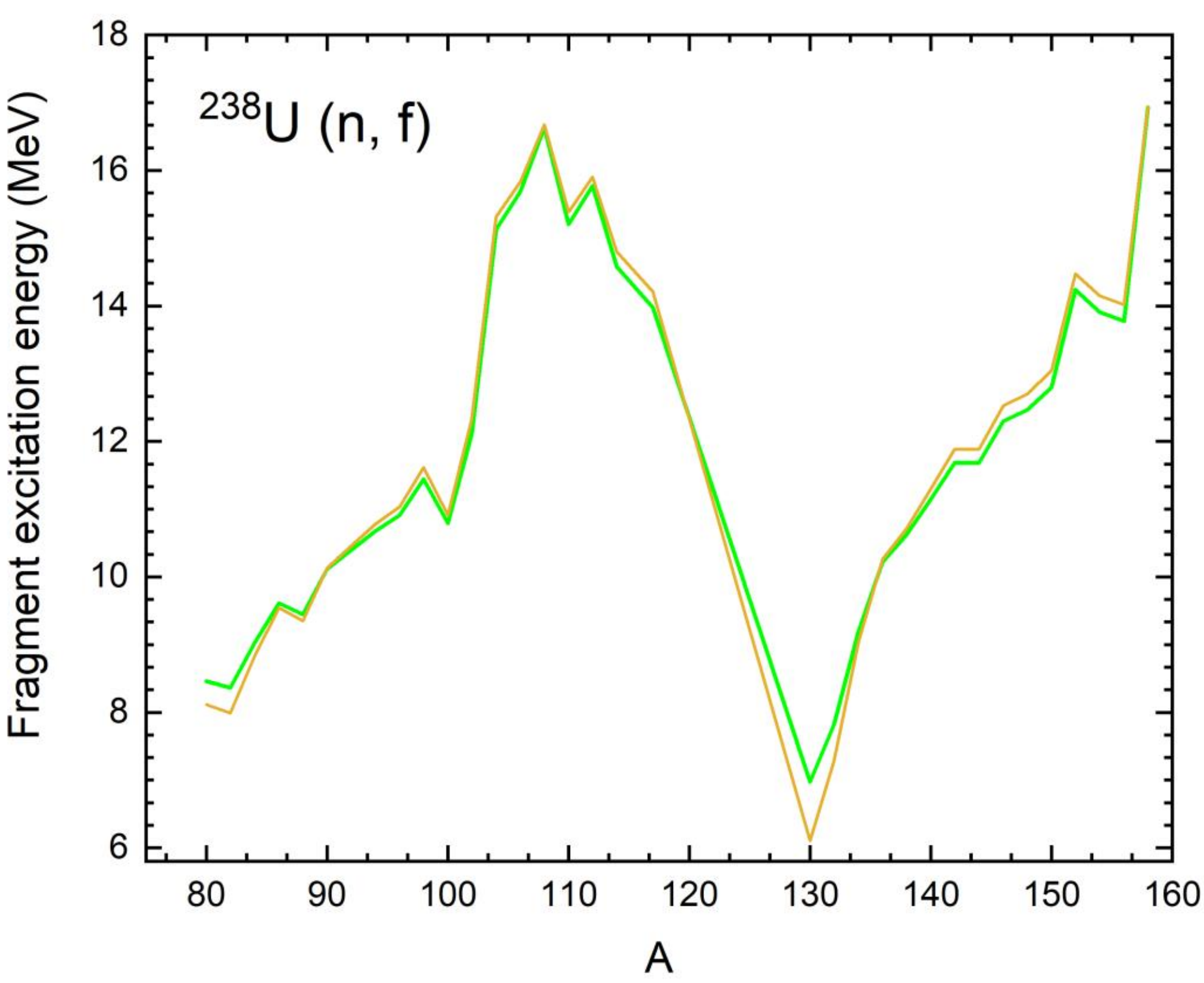

*Fig. 4. Dependence of the average excitation energy on fragment mass in the fission of* $^{238}\mathrm{U}(n,f)$ *by neutrons at threshold energies, calculated using Equation (2) (solid green line) and Equation (3) (yellow line).*

The comparative analysis, presented in Figure 4, demonstrates a high degree of agreement between the two methods based on Equations (2) and (3) across the entire parameter range considered. Both approaches preserve the characteristic sawtooth structure observed in the distribution of neutron multiplicity and the excitation energy behavior of fragments. This structure reflects the complex dynamics of energy redistribution between collective and single-quasiparticle degrees of freedom at different fission stages. Thus, these results confirm the effectiveness of the proposed methods in describing the thermodynamic characteristics of the fission process and their reliability in modeling the dynamic behavior of fission systems.

The next stage of the analysis focuses on establishing the relationship between the excitation energy of fission fragments and their nonequilibrium deformations at the pre-fragmentation stage. To address this, the method proposed by V. Strutinsky in [15] is applied, which utilizes the Nilsson level scheme to calculate shell corrections incorporated into the liquid drop model (LDM) for a more accurate estimation of the total excitation energy.

Strutinsky's corrections depend not only on the number of occupied quantum levels but also on nuclear deformation, enabling a significant improvement in the accuracy of accounting for nuclear degrees of freedom in the total excitation energy. This is particularly critical for analyzing the dynamics of nonequilibrium processes occurring at the pre-fragment formation stage in fission reactions.

In the framework of the considered approach, it is assumed that the total $U$ can be represented through the strain energy calculated within the LDM. This energy is divided into two key components: surface and Coulomb energies. Both of these components can be described using simplified analytical forms, which are defined by Eq. (4):

$$U = \sigma A^{2/3}(0.4(1-x)\alpha^2 - 0.0381(1-2x)\alpha^3), \quad (4)$$

where the coefficients are $\sigma$ = 16 MeV, $x = Z^2/(45A)$, and α the deformation parameter defined by the relation $\alpha = 2\beta/3$.

Using expression (4), we can calculate the equilibrium strain energy by substituting the equilibrium strain values obtained from [23]. Since equations (2) and (3) give a sufficiently high degree of consistency of results with each other for the whole fission fragmentation region, therefore, any of the above formulae can be used in the present work. By adding the excitation energy calculated using either formulae to the equilibrium strain energy, the nonequilibrium excitation energy is obtained.

From the nonequilibrium deformation energy, equation (4) is applied to solve the inverse problem, determining the nonequilibrium deformations of fission pre-fragments. All calculated values are presented in Tables 1 ($^{232}\mathrm{Th}(n,f)$) and 2 (

$^{238}\mathrm{U}(n,f)$).

The structure of Tables 1 and 2 is arranged in a systematic manner. Column 1 lists the fitting parameter d defined in Eq. (6), while Column 2 specifies the corresponding fission fragment pairs. Column 3 presents the quadrupole deformation parameters of the fragments. Columns 4–6 present the ratios of moments of inertia calculated within the oscillator and rectangular well potentials of the superfluid model, as well as the classical hydrodynamic model, normalized to the rigid-body moment of inertia for comparative analysis of pairing quantum effects. Columns 7–8 contain the effective moments of inertia for bending and wriggling vibrations derived from all three nuclear models, whereas Columns 9–11 present the predicted fragment spins calculated on the basis of these models. This arrangement ensures direct comparison between the theoretical approaches while preserving correspondence with experimental observables, in particular for the $^{252}\mathrm{Cf}$ fission system under study. The tabulated data clearly demonstrate how variations in the calculated moment-of-inertia ratios manifest in model-dependent spin distributions, with special emphasis on the sensitivity of the wriggling mode contribution, which had not been explicitly considered in earlier works of this type.

Table1. Quadrupole deformation parameters; moments of inertia (normalized to rigid-body values) for oscillator, rectangular well,
and hydrodynamic models; and resulting fragment spins for $^{232}\mathrm{Th}(n,f)$.

| | Nucleus | $\beta$ | $I_{osc}/I_o$ | $I_{rec}/I_o$ | $I_{hyd}/I_o$ | $I_{b(osc)}$; $I_{b(rec)}$; $I_{b(hyd)}$ | $I_{w(osc)}$; $I_{w(rec)}$; $I_{w(hyd)}$ | $J_{osc}$ | $J_{rec}$ | $J_{hyd}$ |
|---|---|---|---|---|---|---|---|---|---|---|
| d=3.5 | $^{82}$Ge | 0.392 | 0.639 | 0.252 | 0.114 | 62.09<br>34.39<br>19.63 | 76.95<br>40.25<br>22.28 | 5.54 | 4.70 | 4.05 |
| | $^{151}$Ce | 0.654 | 0.854 | 0.473 | 0.270 | | | 8.35 | 7.39 | 6.52 |
| d=4.2 | $^{84}$Ge | 0.418 | 0.688 | 0.320 | 0.129 | 57.71<br>29.77<br>17.26 | 74.57<br>37.60<br>20.42 | 5.46 | 4.59 | 3.92 |
| | $^{149}$Ce | 0.618 | 0.826 | 0.426 | 0.247 | | | 7.85 | 6.72 | 6.03 |
| d=4.1 | $^{84}$Se | 0.415 | 0.595 | 0.228 | 0.127 | 56.22<br>27.76<br>16.27 | 70.77<br>33.34<br>19.38 | 5.34 | 4.41 | 3.85 |
| | $^{149}$Ba | 0.599 | 0.812 | 0.401 | 0.235 | | | 7.82 | 6.69 | 5.88 |
| d=3.4 | $^{86}$Se | 0.437 | 0.699 | 0.324 | 0.139 | 53.04<br>25.18<br>14.56 | 70.99<br>33.51<br>18.13 | 6.10 | 5.05 | 4.30 |
| | $^{147}$Ba | 0.571 | 0.794 | 0.377 | 0.218 | | | 8.57 | 7.13 | 6.40 |
| d=4.4 | $^{88}$Se | 0.482 | 0.728 | 0.334 | 0.164 | 59.83<br>33.07<br>21.72 | 79.66<br>42.17<br>26.19 | 5.95 | 5.04 | 4.45 |
| | $^{145}$Ba | 0.716 | 0.854 | 0.472 | 0.310 | | | 8.52 | 7.48 | 6.91 |
| d=2.9 | $^{88}$Kr | 0.474 | 0.724 | 0.330 | 0.160 | 52.15<br>25.10<br>13.33 | 71.80<br>34.06<br>17.67 | 6.79 | 5.62 | 4.75 |
| | $^{145}$Xe | 0.552 | 0.806 | 0.388 | 0.206 | | | 9.38 | 7.86 | 6.77 |
| d=2.85 | $^{90}$Kr | 0.496 | 0.720 | 0.314 | 0.173 | 49.21<br>22.36<br>11.87 | 69.71<br>31.30<br>16.79 | 6.74 | 5.51 | 4.72 |
| | $^{143}$Xe | 0.526 | 0.788 | 0.358 | 0.190 | | | 9.09 | 7.50 | 6.37 |
| d=2.6 | $^{92}$Kr | 0.520 | 0.737 | 0.319 | 0.187 | 48.78<br>22.56<br>11.59 | 70.79<br>32.09<br>17.17 | 7.22 | 5.90 | 5.09 |
| | $^{141}$Xe | 0.525 | 0.800 | 0.370 | 0.190 | | | 9.58 | 7.97 | 6.63 |
| d=2.05 | $^{94}$Kr | 0.550 | 0.739 | 0.320 | 0.205 | 47.55<br>21.99<br>11.11 | 70.74<br>32.03<br>17.55 | 8.40 | 6.86 | 6.03 |
| | $^{139}$Xe | 0.521 | 0.800 | 0.370 | 0.187 | | | 11.01 | 9.16 | 7.48 |
| d=3.3 | $^{92}$Sr | 0.518 | 0.786 | 0.373 | 0.186 | 47.63<br>22.59<br>10.01 | 71.08<br>33.71<br>15.55 | 5.88 | 4.88 | 4.06 |
| | $^{141}$Te | 0.487 | 0.795 | 0.377 | 0.167 | | | 7.54 | 6.25 | 5.02 |
| d=3.1 | $^{94}$Sr | 0.539 | 0.765 | 0.361 | 0.198 | 47.48<br>22.86<br>10.37 | 71.37<br>34.13<br>16.55 | 6.35 | 5.28 | 4.48 |
| | $^{139}$Te | 0.502 | 0.806 | 0.388 | 0.176 | | | 8.14 | 6.80 | 5.44 |
| d=2.6 | $^{96}$Sr | 0.649 | 0.806 | 0.390 | 0.267 | 44.42<br>20.42<br>7.70 | 71.89<br>33.72<br>16.80 | 7.30 | 6.07 | 5.55 |
| | $^{137}$Te | 0.436 | 0.796 | 0.366 | 0.138 | | | 8.84 | 7.23 | 5.15 |
| d=2.7 | $^{98}$Sr | 0.719 | 0.869 | 0.488 | 0.312 | 43.91<br>20.31<br>8.12 | 75.65<br>38.13<br>19.52 | 7.72 | 6.69 | 6.10 |
| | $^{135}$Te | 0.454 | 0.800 | 0.370 | 0.148 | | | 8.87 | 7.09 | 5.20 |
| d=2.55 | $^{98}$Zr | 0.731 | 0.792 | 0.380 | 0.320 | 38.35<br>18.57<br>7.57 | 67.44<br>32.54<br>19.33 | 7.79 | 6.49 | 6.41 |
| | $^{135}$Sn | 0.437 | 0.704 | 0.341 | 0.139 | | | 8.80 | 7.36 | 5.19 |
| d=2.9 | $^{100}$Zr | 0.735 | 0.854 | 0.454 | 0.323 | 38.36<br>19.71<br>7.57 | 70.87<br>37.00<br>19.86 | 7.46 | 6.37 | 6.03 |
| | $^{133}$Sn | 0.443 | 0.720 | 0.370 | 0.142 | | | 8.03 | 6.76 | 4.79 |
| d=3.9 | $^{130}$Sn | 0.411 | 0.583 | 0.230 | 0.125 | 29.48<br>11.63<br>6.32 | 66.14<br>33.36<br>20.47 | 5.25 | 3.89 | 3.25 |
| | $^{103}$Zr | 0.774 | 0.899 | 0.533 | 0.347 | | | 5.80 | 5.21 | 4.74 |
| d=3.7 | $^{132}$Sn | 0.358 | 0.388 | 0.155 | 0.098 | 19.66<br>7.85<br>4.97 | 53.24<br>25.70<br>18.41 | 5.49 | 4.18 | 3.64 |
| | $^{101}$Zr | 0.766 | 0.854 | 0.454 | 0.342 | | | 7.02 | 6.11 | 5.76 |
| d=4.1 | $^{132}$Te | 0.371 | 0.682 | 0.264 | 0.104 | 34.76<br>13.45<br>5.30 | 69.53<br>33.17<br>17.95 | 6.23 | 4.62 | 3.40 |
| | $^{101}$Sr | 0.740 | 0.896 | 0.508 | 0.326 | | | 6.23 | 5.48 | 5.05 |
| d=4.6 | $^{134}$Te | 0.420 | 0.599 | 0.237 | 0.130 | 31.98<br>12.65<br>6.94 | 66.07<br>31.97<br>18.80 | 5.47 | 4.10 | 3.46 |
| | $^{99}$Sr | 0.727 | 0.914 | 0.518 | 0.318 | | | 5.62 | 4.94 | 4.40 |
| d=3.2 | $^{136}$Te | 0.430 | 0.738 | 0.320 | 0.135 | 40.57<br>17.59<br>7.42 | 70.29<br>33.26<br>17.53 | 8.08 | 6.36 | 4.78 |
| | $^{97}$Sr | 0.680 | 0.844 | 0.445 | 0.287 | | | 7.11 | 6.05 | 5.49 |
| d=3.4 | $^{138}$Te | 0.455 | 0.739 | 0.321 | 0.149 | 42.10<br>18.29<br>8.49 | 70.40<br>33.73<br>16.87 | 7.53 | 5.90 | 4.75 |
| | $^{95}$Sr | 0.627 | 0.854 | 0.466 | 0.253 | | | 6.44 | 5.51 | 4.73 |
| d=3.8 | $^{138}$Xe | 0.473 | 0.737 | 0.319 | 0.159 | 42.33<br>18.32<br>9.13 | 68.36<br>30.53<br>17.40 | 7.01 | 5.62 | 4.51 |
| | $^{95}$Kr | 0.623 | 0.787 | 0.369 | 0.250 | | | 5.83 | 4.80 | 4.33 |
| d=2.2 | $^{140}$Xe | 0.516 | 0.749 | 0.327 | 0.184 | 44.94<br>19.62<br>11.04 | 69.03<br>30.94<br>17.58 | 11.12 | 8.95 | 7.72 |
| | $^{93}$Kr | 0.561 | 0.777 | 0.365 | 0.211 | | | 8.75 | 7.21 | 6.28 |
| d=2.2 | $^{142}$Xe | 0.517 | 0.765 | 0.335 | 0.185 | 47.02<br>20.59<br>11.37 | 69.83<br>30.83<br>16.72 | 10.90 | 8.84 | 7.67 |
| | $^{91}$Kr | 0.514 | 0.780 | 0.350 | 0.183 | | | 8.36 | 6.83 | 5.83 |
| d=3.3 | $^{142}$Ba | 0.535 | 0.754 | 0.344 | 0.196 | 46.74<br>21.32<br>12.15 | 68.95<br>32.57<br>17.37 | 8.13 | 6.59 | 5.88 |
| | $^{91}$Se | 0.507 | 0.762 | 0.386 | 0.179 | | | 6.26 | 5.23 | 4.40 |
| d=2.7 | $^{144}$Ba | 0.558 | 0.777 | 0.366 | 0.210 | 49.84<br>23.48<br>13.47 | 71.25<br>33.65<br>18.18 | 10.02 | 8.29 | 7.40 |
| | $^{89}$Se | 0.491 | 0.768 | 0.365 | 0.169 | | | 7.45 | 6.18 | 5.24 |
| d=3.0 | $^{146}$Ba | 0.565 | 0.771 | 0.353 | 0.214 | 50.77<br>23.25<br>14.09 | 70.98<br>32.08<br>18.09 | 9.37 | 7.75 | 7.05 |
| | $^{87}$Se | 0.459 | 0.764 | 0.334 | 0.151 | | | 6.88 | 5.62 | 4.82 |
| d=3.0 | $^{148}$Ce | 0.594 | 0.772 | 0.355 | 0.232 | 52.73<br>24.25<br>15.85 | 79.08<br>33.46<br>19.32 | 9.85 | 8.08 | 7.67 |
| | $^{85}$Ge | 0.435 | 0.769 | 0.366 | 0.138 | | | 7.08 | 5.86 | 5.02 |
| d=2.9 | $^{150}$Ce | 0.640 | 0.840 | 0.448 | 0.261 | 59.99<br>31.99<br>18.64 | 77.99<br>39.87<br>21.71 | 11.22 | 9.77 | 7.80 |
| | $^{83}$Ge | 0.640 | 0.750 | 0.328 | 0.128 | | | 7.67 | 6.45 | 5.52 |

Table2. Quadrupole deformation parameters; moments of inertia (normalized to rigid-body values) for oscillator, rectangular well,
and hydrodynamic models; and resulting fragment spins for $^{238}U(n,f)$.

| | Nucle us | $\beta$ | $I_{osc}/I_o$ | $I_{rec}/I_o$ | $I_{hyd}/I_o$ | $I_{b(osc)}$;<br>$I_{b(rec)}$;<br>$I_{b(hyd)}$ | $I_{w(osc)}$;<br>$I_{w(rec)}$;<br>$I_{w(hyd)}$ | $J_{osc}$ | $J_{rec}$ | $J_{hyd}$ |
|---|---|---|---|---|---|---|---|---|---|---|
| d=4.7 | $^{82}$Ge | 0.372 | 0.599 | 0.232 | 0.105 | 72.80<br>43.10<br>25.28 | 86.61<br>48.45<br>27.07 | 5.92 | 5.12 | 4.46 |
| | $^{157}$Nd | 0.723 | 0.907 | 0.537 | 0.315 | | | 9.15 | 8.24 | 7.30 |
| d=4.5 | $^{84}$Se | 0.516 | 0.717 | 0.295 | 0.184 | 62.13<br>31.40<br>18.69 | 80.50<br>38.96<br>23.40 | 5.43 | 4.50 | 3.97 |
| | $^{155}$Ce | 0.622 | 0.831 | 0.420 | 0.250 | | | 7.85 | 6.74 | 5.90 |
| d=4.1 | $^{86}$Se | 0.527 | 0.766 | 0.366 | 0.191 | 64.14<br>34.24<br>22.90 | 84.65<br>44.04<br>28.02 | 6.49 | 5.48 | 4.87 |
| | $^{153}$Ce | 0.702 | 0.843 | 0.450 | 0.301 | | | 9.41 | 8.13 | 7.52 |
| d=4.1 | $^{88}$Se | 0.536 | 0.756 | 0.356 | 0.197 | 63.27<br>35.07<br>21.76 | 84.39<br>45.02<br>27.26 | 6.37 | 5.41 | 4.76 |
| | $^{151}$Ce | 0.690 | 0.855 | 0.474 | 0.294 | | | 9.15 | 8.03 | 7.20 |
| d=4.3 | $^{88}$Kr | 0.533 | 0.770 | 0.366 | 0.194 | 61.41<br>32.86<br>20.15 | 82.89<br>43.07<br>25.67 | 5.96 | 5.03 | 4.40 |
| | $^{151}$Ba | 0.663 | 0.841 | 0.450 | 0.276 | | | 8.41 | 7.29 | 6.54 |
| d=3.9 | $^{90}$Kr | 0.561 | 0.772 | 0.368 | 0.211 | 58.76<br>29.45<br>19.54 | 81.42<br>40.25<br>25.73 | 6.45 | 5.39 | 4.79 |
| | $^{149}$Ba | 0.660 | 0.824 | 0.413 | 0.274 | | | 8.97 | 7.59 | 6.96 |
| d=3.5 | $^{92}$Kr | 0.602 | 0.783 | 0.371 | 0.237 | 56.76<br>28.24<br>18.53 | 81.06<br>39.76<br>25.88 | 6.94 | 5.78 | 5.19 |
| | $^{147}$Ba | 0.650 | 0.818 | 0.407 | 0.267 | | | 9.46 | 8.00 | 7.22 |
| d=3.5 | $^{94}$Kr | 0.632 | 0.784 | 0.372 | 0.256 | 57.01<br>31.07<br>17.52 | 82.61<br>43.21<br>25.88 | 6.97 | 5.86 | 5.24 |
| | $^{145}$Ba | 0.637 | 0.846 | 0.461 | 0.260 | | | 9.40 | 8.22 | 6.94 |
| d=3.7 | $^{94}$Sr | 0.627 | 0.811 | 0.407 | 0.253 | 53.44<br>26.39<br>15.01 | 79.86<br>39.65<br>23.25 | 6.40 | 5.38 | 4.74 |
| | $^{145}$Xe | 0.588 | 0.812 | 0.401 | 0.228 | | | 8.33 | 6.97 | 5.97 |
| d=3.45 | $^{96}Sr$ | 0.659 | 0.816 | 0.416 | 0.273 | 53.02<br>26.96<br>14.96 | 80.97<br>41.21<br>24.32 | 6.78 | 5.72 | 5.11 |
| | $^{143}Xe$ | 0.594 | 0.822 | 0.418 | 0.232 | | | 8.70 | 7.34 | 6.18 |
| d=2.8 | $^{98}Sr$ | 0.676 | 0.861 | 0.480 | 0.284 | 51.61<br>26.25<br>14.32 | 82.40<br>43.41<br>24.47 | 8.14 | 7.01 | 6.14 |
| | $^{141}Xe$ | 0.587 | 0.822 | 0.418 | 0.228 | | | 10.09 | 8.40 | 7.12 |
| d=2.9 | $^{98}Zr$ | 0.688 | 0.770 | 0.366 | 0.292 | 51.40<br>26.38<br>13.22 | 79.09<br>39.55<br>23.72 | 7.70 | 6.42 | 5.98 |
| | $^{141}Te$ | 0.563 | 0.828 | 0.425 | 0.213 | | | 9.88 | 8.45 | 6.62 |
| d=2.9 | $^{100}Zr$ | 0.686 | 0.830 | 0.436 | 0.291 | 49.40<br>24.58<br>12.56 | 80.24<br>40.78<br>23.37 | 7.85 | 6.68 | 6.03 |
| | $^{139}Te$ | 0.555 | 0.818 | 0.407 | 0.208 | | | 9.55 | 7.96 | 6.44 |
| d=3.0 | $^{102}Zr$ | 0.717 | 0.842 | 0.449 | 0.311 | 48.84<br>25.36<br>11.86 | 81.67<br>42.87<br>23.99 | 7.86 | 6.71 | 6.16 |
| | $^{137}Te$ | 0.546 | 0.832 | 0.432 | 0.202 | | | 9.31 | 7.87 | 6.09 |
| d=3.5 | $^{104}Zr$ | 0.754 | 0.849 | 0.462 | 0.334 | 47.75<br>25.01<br>10.91 | 82.59<br>43.97<br>24.62 | 7.16 | 6.14 | 5.75 |
| | $^{135}Te$ | 0.530 | 0.840 | 0.440 | 0.192 | | | 8.20 | 6.93 | 5.17 |
| d=2.0 | $^{102}Mo$ | 0.712 | 0.797 | 0.384 | 0.308 | 45.54<br>23.77<br>12.06 | 76.54<br>38.71<br>24.04 | 9.73 | 8.11 | 7.74 |
| | $^{133}Sn$ | 0.551 | 0.774 | 0.404 | 0.205 | | | 11.56 | 9.95 | 7.77 |

| | | | | | | | | | |
|---|---|---|---|---|---|---|---|---|---|
| d=3.1 | $^{104}$Mo | 0.776 | 0.841 | 0.447 | 0.349 | 31.31<br>12.22<br>10.14 | 66.20<br>30.77<br>24.62 | 7.99 | 6.95 | 6.51 |
| | $^{131}$Sn | 0.510 | 0.556 | 0.217 | 0.180 | | | 7.59 | 5.69 | 5.48 |
| d=5.0 | $^{130}$Sn | 0.452 | 0.620 | 0.250 | 0.147 | 31.93<br>12.87<br>7.57 | 69.81<br>37.95<br>22.55 | 5.44 | 4.01 | 3.51 |
| | $^{109}$Mo | 0.758 | 0.852 | 0.464 | 0.337 | | | 5.86 | 5.42 | 4.77 |
| d=4.9 | $^{132}$Sn | 0.470 | 0.535 | 0.207 | 0.157 | 28.49<br>11.02<br>8.36 | 65.06<br>30.99<br>22.73 | 5.34 | 3.99 | 3.75 |
| | $^{107}$Mo | 0.754 | 0.850 | 0.464 | 0.334 | | | 5.96 | 5.21 | 4.79 |
| d=4.7 | $^{134}$Sn | 0.503 | 0.728 | 0.384 | 0.176 | 40.36<br>21.29<br>9.76 | 76.47<br>41.00<br>24.96 | 6.61 | 5.60 | 4.23 |
| | $^{105}$Mo | 0.791 | 0.850 | 0.464 | 0.358 | | | 6.30 | 5.42 | 5.19 |
| | $^{132}$Te | 0.487 | 0.770 | 0.336 | 0.167 | 41.33<br>18.03<br>8.96 | 79.33<br>40.56<br>22.90 | 5.98 | 4.63 | 3.75 |
| | $^{103}$Zr | 0.741 | 0.889 | 0.527 | 0.326 | | | 5.77 | 5.10 | 4.57 |
| d=4.8 | $^{134}$Te | 0.522 | 0.708 | 0.291 | 0.188 | 39.60<br>16.28<br>10.52 | 74.76<br>34.98<br>24.55 | 6.29 | 4.83 | 4.22 |
| | $^{101}$Zr | 0.757 | 0.842 | 0.448 | 0.336 | | | 5.98 | 5.13 | 4.81 |
| d=3.4 | $^{136}$Te | 0.536 | 0.801 | 0.386 | 0.196 | 46.23<br>22.28<br>11.31 | 79.75<br>40.14<br>24.14 | 8.98 | 7.36 | 5.86 |
| | $^{99}$Zr | 0.733 | 0.839 | 0.447 | 0.321 | | | 7.83 | 6.68 | 6.21 |

| | | | | | | | | | |
|---|---|---|---|---|---|---|---|---|---|
| d=3.7 | $^{138}$Te | 0.543 | 0.796 | 0.381 | 0.200 | 47.23<br>22.60<br>11.87 | 80.58<br>41.15<br>23.33 | 8.26 | 6.71 | 5.56 |
| | $^{97}$Zr | 0.702 | 0.876 | 0.487 | 0.301 | | | 7.15 | 6.16 | 5.48 |
| d=3.8 | $^{138}$Xe | 0.565 | 0.794 | 0.379 | 0.214 | 47.60<br>22.72<br>12.83 | 80.32<br>41.09<br>23.60 | 8.15 | 6.60 | 5.69 |
| | $^{101}$Sr | 0.679 | 0.869 | 0.488 | 0.286 | | | 6.99 | 6.03 | 5.28 |
| d=2.8 | $^{140}$Xe | 0.582 | 0.783 | 0.367 | 0.224 | 48.47<br>22.72<br>13.87 | 80.01<br>40.39<br>24.06 | 10.39 | 8.35 | 7.45 |
| | $^{99}$Sr | 0.672 | 0.869 | 0.487 | 0.281 | | | 8.66 | 7.48 | 6.52 |
| d=2.9 | $^{142}$Xe | 0.593 | 0.799 | 0.388 | 0.231 | 50.91<br>24.72<br>14.72 | 81.48<br>41.90<br>24.57 | 10.45 | 8.53 | 7.54 |
| | $^{97}$Sr | 0.671 | 0.872 | 0.490 | 0.281 | | | 8.46 | 7.30 | 3.56 |
| d=3.4 | $^{142}$Ba | 0.616 | 0.794 | 0.378 | 0.246 | 51.15<br>24.35<br>15.85 | 79.53<br>38.19<br>25.11 | 9.48 | 7.85 | 7.02 |
| | $^{97}$Kr | 0.649 | 0.818 | 0.399 | 0.267 | | | 7.50 | 6.26 | 5.66 |
| d=3.0 | $^{144}$Ba | 0.620 | 0.721 | 0.387 | 0.248 | 47.64<br>25.57<br>16.39 | 74.08<br>38.31<br>25.05 | 10.18 | 8.85 | 7.85 |
| | $^{95}$Kr | 0.639 | 0.793 | 0.382 | 0.260 | | | 8.03 | 6.74 | 6.10 |
| d=3.2 | $^{146}$Ba | 0.620 | 0.795 | 0.380 | 0.248 | 53.75<br>25.69<br>16.77 | 78.76<br>37.61<br>24.03 | 10.02 | 8.33 | 7.56 |
| | $^{93}$Kr | 0.592 | 0.795 | 0.379 | 0.231 | | | 7.53 | 6.25 | 5.56 |
| d=4.3 | $^{148}$Ba | 0.659 | 0.785 | 0.370 | 0.248 | 55.33<br>26.08<br>17.48 | 79.88<br>38.20<br>24.21 | 8.10 | 6.67 | 6.18 |
| | $^{91}$Kr | 0.580 | 0.814 | 0.402 | 0.223 | | | 6.06 | 5.06 | 4.45 |
| d=4.0 | $^{148}$Ce | 0.680 | 0.794 | 0.379 | 0.287 | 56.54<br>26.99<br>20.44 | 80.40<br>39.29<br>26.92 | 8.97 | 7.38 | 7.19 |
| | $^{91}$Se | 0.569 | 0.795 | 0.410 | 0.216 | | | 6.58 | 5.54 | 4.93 |
| d=3.1 | $^{150}$Ce | 0.684 | 0.842 | 0.449 | 0.289 | 61.44<br>32.76<br>21.09 | 84.61<br>44.67<br>27.00 | 11.52 | 9.89 | 9.11 |
| | $^{89}$Se | 0.552 | 0.808 | 0.415 | 0.206 | | | 8.13 | 6.91 | 6.02 |

## C. Finding the optimal model

In this study, calculations of the moments of inertia of fission fragments for the nuclei $^{232}\mathrm{Th}(n,f)$ and $^{238}\mathrm{U}(n,f)$ were performed on the basis of the values of nonequilibrium quadrupole strains presented in Tables I and II. These calculations employed the methodology proposed in [18], which combines hydrodynamic [24] and superfluid models, incorporating oscillatory and rectangular potentials. This approach enables more comprehensive and accurate modeling of the dynamics of nuclear fission processes, providing a detailed description of various fission stages.

The resulting values of the nonequilibrium moments of inertia are presented in Tables 1 and 2, offering important information for further analyses of the dynamical aspects of nuclear fission.

However, the lack of direct experimental data that could serve as a criterion for verifying the theoretical models requires the use of indirect methods to validate the calculations. The study uses the approach proposed in [18], which consists in comparing the theoretical values of the mean spins of fission fragments with the experimental data published in [1], obtained for forced threshold fission of nuclei $^{232}\mathrm{Th}(n,f)$ and $^{238}\mathrm{U}(n,f)$. This method provides an additional validation of the theoretical models' accuracy by comparing them with experimentally measured values, serving as an important step in refining the models and enhancing their predictive capabilities.

To perform the calculations, this work utilizes a formalism developed in recent studies [25]. Considering that fission fragments emerging from the CFS reach the region near the scission point exclusively in cold nonequilibrium states [11], the calculation of their mean spin values is conducted based on zero-point oscillatory wave functions in the momentum representation, as part of a more general approach [9, 25].

$$\Psi\left(J_{k_x}, J_{k_y}\right) \equiv \Psi\left(J_{k_x}\right)\Psi\left(J_{k_y}\right) = \frac{1}{\pi I_k \hbar\omega_k}\exp\left[-\frac{J_{k_x}^2 + J_{k_y}^2}{I_k \hbar\omega_k}\right], \quad (5)$$

where the index $k$ denotes the type of oscillations bending ($b$) or wriggling($w$), the energies and moments of inertia of the specified zero oscillations $\hbar\omega_w$ and $I_w$, $\hbar\omega_b$ and $I_b$ respectively. Determining the moments of inertia corresponding to transverse bending and wriggling oscillations for each particular pair of primary fission fragment (PFF) pre-delimitation formations is a challenging theoretical problem. In particular, for wiggling oscillations, the moment of inertia $I_w$, as demonstrated in [26, 27], can be calculated using the following formula:

$$I_w = \frac{(I_1 + I_2) I_0}{I}, \quad (6)$$

where $I_1$ $(I_2)$ are the moments of inertia of each fragment, $I_0 = \frac{M_1 M_2}{(M_1 + M_2)} (R_1 + R_2 + d)^2$ is the moment of inertia of the fission core, $d$ – is a fitting parameter representing the distance between the ends of the fission pre-fragments; the short-range character of nuclear forces causes even small distance variations (0.5-1 fm) to dramatically alter the stiffness of both bending and wriggling vibrations, resulting in a two-fold change in spin values. Since parameter «d» strongly affects the nuclear component of the compound nucleus potential, its variation requires careful optimization. For all three models considered, we have determined the optimal «d»-value that provides the best agreement with experimental spin data; $I = I_0 + I_1 + I_2$ is the total moment of inertia.

Moments of inertia in the solid state model, can be represented by: $I_{1,2} \equiv I_{i,\mathrm{rigid}} = \frac{M_i}{5} \sum R_i^2$, $M_i$ is the mass of each fission fragment, $R_i = r_0 A^{1/3} \left[ 1 - \beta_i^2 / 4\pi + \beta_i \sqrt{5/4\pi} \right]$, $\beta_i$ is the quadrupole deformation parameter, with $I_{1,2}$ determined within us in the framework of the superfluid Migdal nucleus model [28] and differing significantly from their rigid-body counterparts $I_i = (0.4 - 0.7) I_{\mathrm{rigit}}$, particularly for fragments near "magic" nuclei, where $I_i = (0.2 - 0.3) I_{\mathrm{rigid}}$.

In this study, the formula proposed in [29] is applied for the moment of inertia $I_b$ :

$$I_b = \mu R^2 I_H / \left( \mu R^2 + I_H \right). \quad (7)$$

where $\mu = M1M2/(M1+M2)$ is the reduced mass and $I_H$ is the moment of inertia of the heavy fragment.

Using this form of SDs and performing several simple transformations, expressions for calculating the mean PFF spins can be obtained:

$$\bar{J}_i = \int_0^{\infty} P(J_i)\, J_i\, dJ_i = \int_0^{\infty} \frac{2J_i^2}{d_i} \exp\left[-\frac{J_i^2}{d_i}\right] dJ_i = \frac{1}{2}\sqrt{\pi d_i}\,. \quad (8)$$

where $d_i = \frac{I_i^2 I_w \hbar\omega_w}{(I_1+I_2)^2} + I_b \hbar\omega_b$, and $i$=(1,2) indexes the corresponding fragment. Subsequently, using formula (8), estimates of spins for three different models of moments of inertia are obtained, as presented in Figs. 5 and 6 for the nuclei $^{232}\mathrm{Th}(n,f)$ and $^{238}\mathrm{U}(n,f)$, respectively.

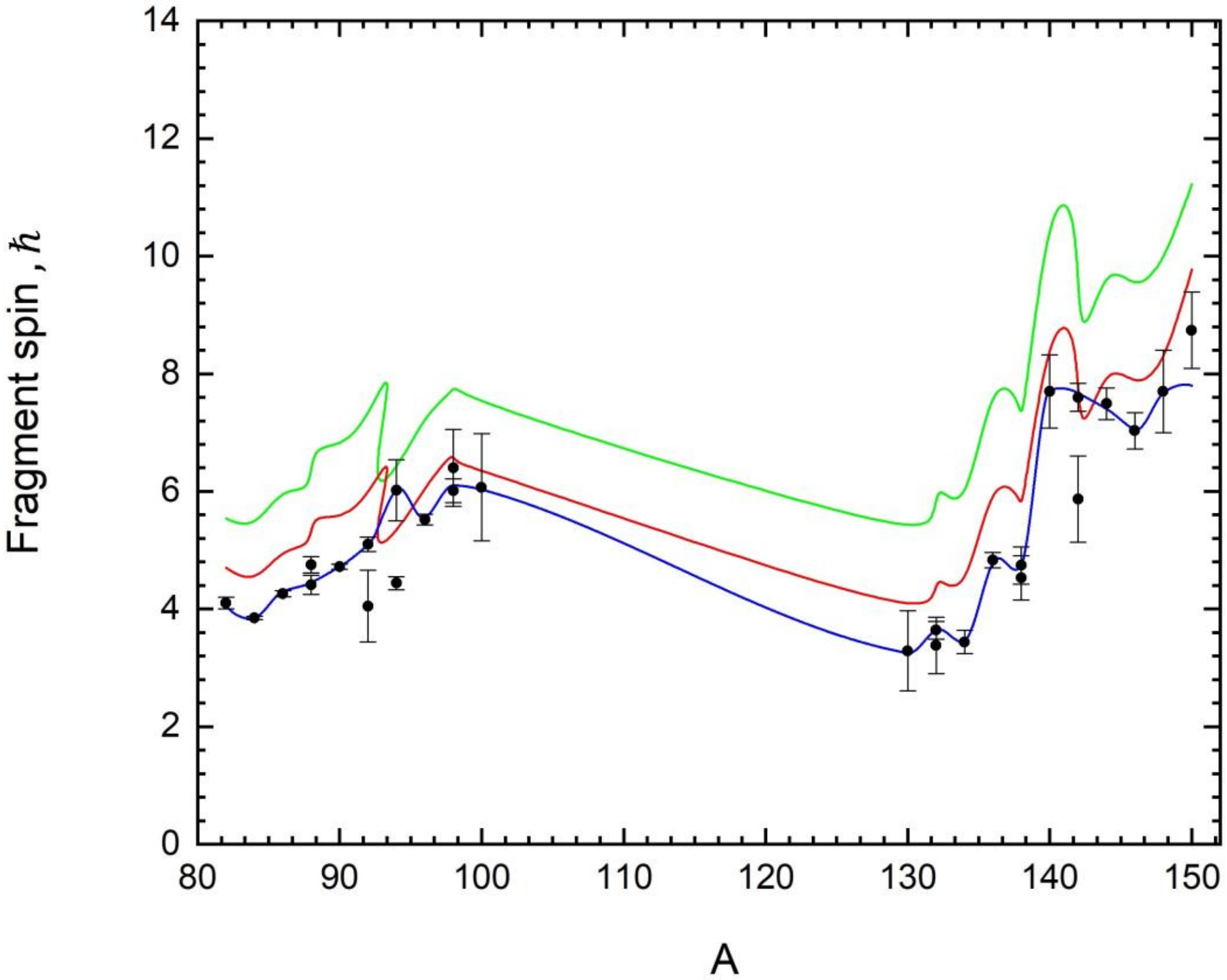

*Fig. 5. Dependence of the mean spin on the mass of forced threshold fission fragments of* $^{232}\mathrm{Th}$, *obtained using three different estimates of the moments of*

*inertia. The green line corresponds to the superfluid approach with an oscillatory potential, the red line to the rectangular potential, and the blue line to the hydrodynamic model. Experimental data are from [1].*

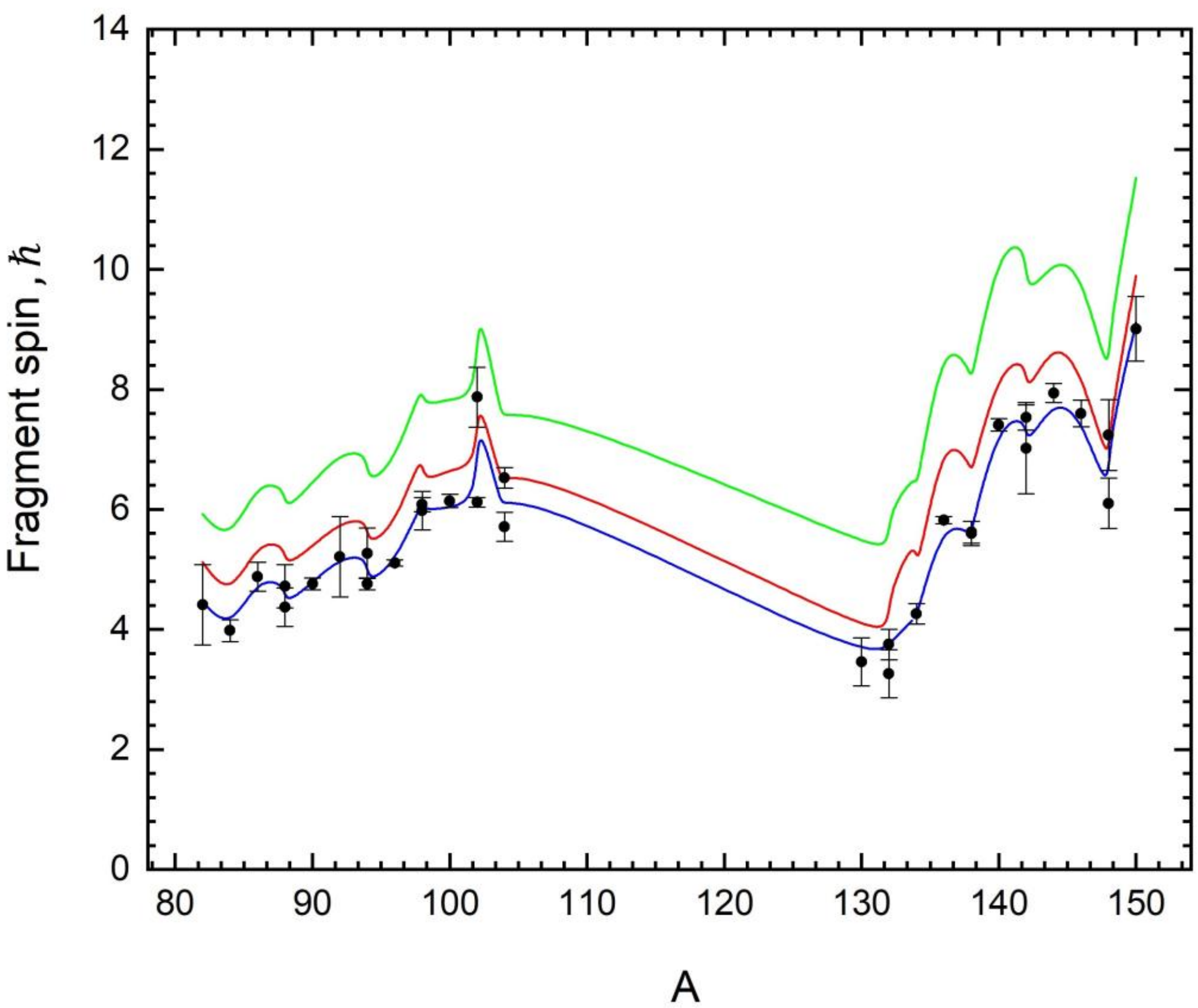

*Fig. 6. Dependence of the mean spin on the mass of forced threshold fission fragments of* $^{238}\mathrm{U}$*, calculated using three different estimates of the moments of inertia. The green line corresponds to the superfluid approach with an oscillatory potential, the red line to the rectangular potential, and the blue line to the hydrodynamic model. Experimental data are from [1].*

A comparative analysis of the theoretical curves with the experimental data presented in [1] shows reasonable agreement for both nuclei in the entire region of light and heavy fragments with mass numbers from 80 to 150, where the moments of inertia are estimated within the hydrodynamic approach. In [18], the moments of inertia were evaluated for the entire working region of fission fragments in the spontaneous fission of the nucleus $^{252}\mathrm{Cf}$.

The study found that, for the entire specified region, the hydrodynamic model is most applicable because of strong nonequilibrium quadrupole deformations significantly exceeding the equilibrium v alues. In such cases, the nucleon spacings become smaller than the nucleus size, leading to the predominance of collective effects. In such cases, the nucle on spacings become smaller than the nucleus size, leading to the predominance of collective effects. In the context of threshold fission of the considered nuclei $^{232}\mathrm{Th}(n,f)$ and $^{238}\mathrm{U}(n,f)$, a similar relation between equilibrium and nonequilibrium deformation is preserved as seen in the whole region, therefore, as in the case of spontaneous fission, the most accurate agreement is achieved using the hydrodynamic model. An additional factor contributing to this agreement was the consideration of neck dynamics between fragments. These results highlight the need for a deeper theory capable of integrating the mechanisms of formation of moments of inertia, vibrational frequencies and fragment spins, addressing broader aspects of nuclear fission. Given these difficulties, the choice to focus on the forced threshold fission of $^{232}\mathrm{Th}(n,f)$ and $^{238}\mathrm{U}(n,f)$ nuclei were justified by the need to work with the most complete and reliable experimental data. The next section details why these particular isotopes were chosen and discuss the limitations of data availability for the other nuclei.

## DISCUSSION

The study was originally planned within a broader context, encompassing a range of actinide isotopes undergoing forced fission by threshold and thermal energy neutrons, such as $^{232}\mathrm{Th}$, $^{238-242}\mathrm{Pu}$, $^{233-238}\mathrm{U}$, to perform a comprehensive comparative analysis of the fission mechanism of these nuclei. However, the study faced significant limitations owing to the scarcity of experimental data. To perform calculations based on the proposed methodology [15], we require experimental values of the instantaneous neutron multiplicity. This constraint considerably limited the scope of the study, focusing only isotopes such as $^{232}\mathrm{Th}$ [30], $^{238}\mathrm{U}$ [17], $^{240}\mathrm{Pu}$ [31], $^{235}U$, $^{237}Np$ and $^{239}Pu$ [32], for which appropriate experimental data are available.

Additionally, verification and validation of the computational results require similar experimental neutron
distributions for other isotopes, which posed a significant
obstacle. The lack of such data for several other actinide
nuclei has significantly restricted the possibility of conducting comprehensive mod el verification and broader comparative analysis.

Another important aspect is the availability of data on SDs for the aforementioned actinide nuclei. In addition to the results discussed earlier for the nuclei $^{232}Th$ and $^{238}U$, described in [1], similar experimental SD data for heavy fission fragments were found in [31]. However, these data raise significant doubts. The main problem is that the values of spins in this study appear to be overestimated. This is clearly demonstrated in Figures 7 and 8, where a visual comparison is presented. A particularly noticeable overestimation of spin values is observed in the interval of mass numbers $128 \leq A_f \leq 140$, which attracts attention because this region is close to magic nuclei. Moreover, this is particularly pronounced for even-even nuclei in this region, whereas for even-even nuclei, the deviations are less significant. This discrepancy is evident since the fragments in this region should have small moments of inertia, according to Tables I ($^{232}Th(n,f)$) and II ($^{238}U(n,f)$), which determine the formation of spins, as correctly noted in [2, 3]. The observed discrepancy for both nuclei underscores the importance of using only modern data on the SD of fission fragments for a correct comparison.

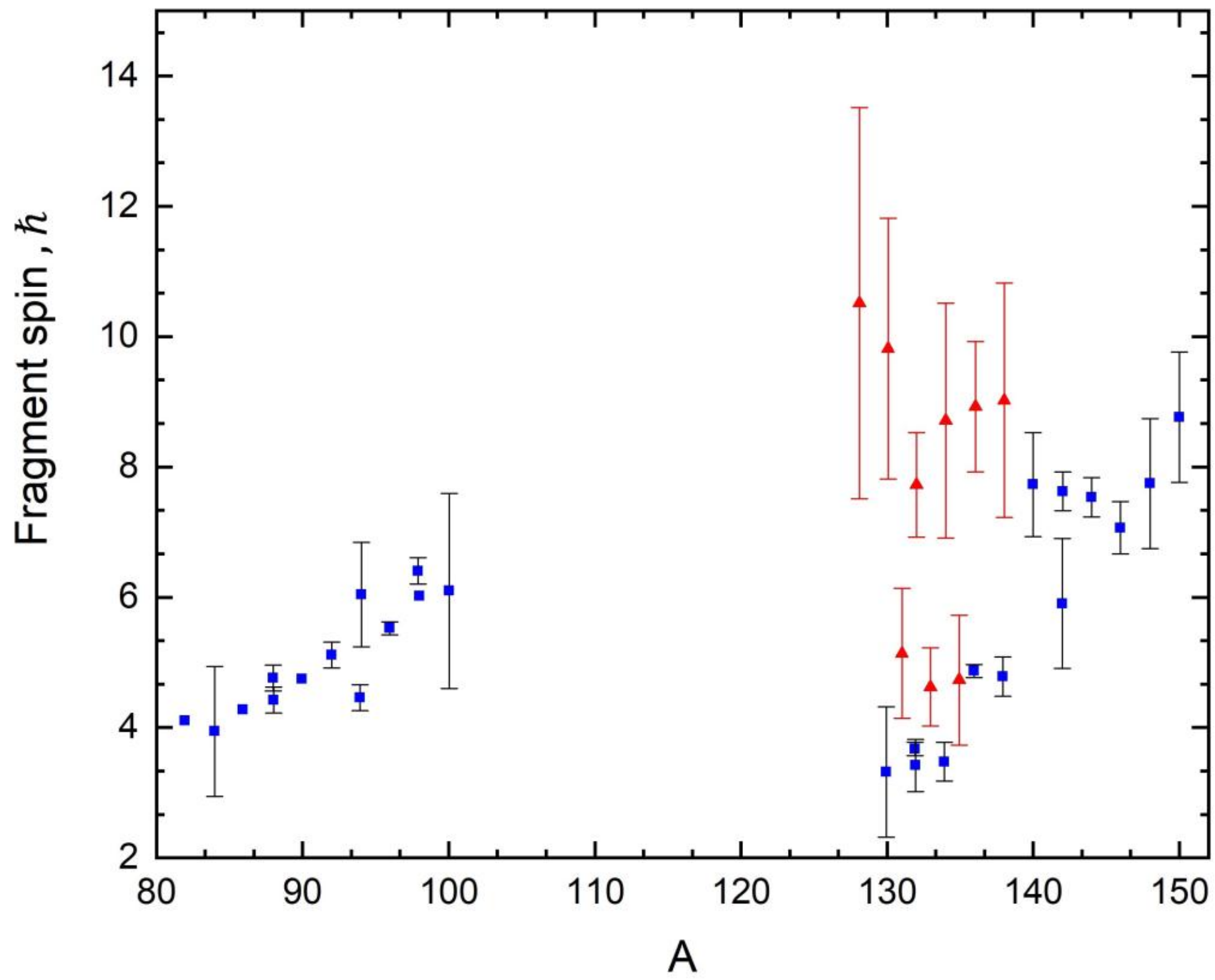

*Fig. 7. Comparison of experimental average spin values as a function of the mass of forced fission fragments of* $^{232}\mathrm{Th}(n,f)$*. Red triangles indicate results from [31], and blue squares from [1].*

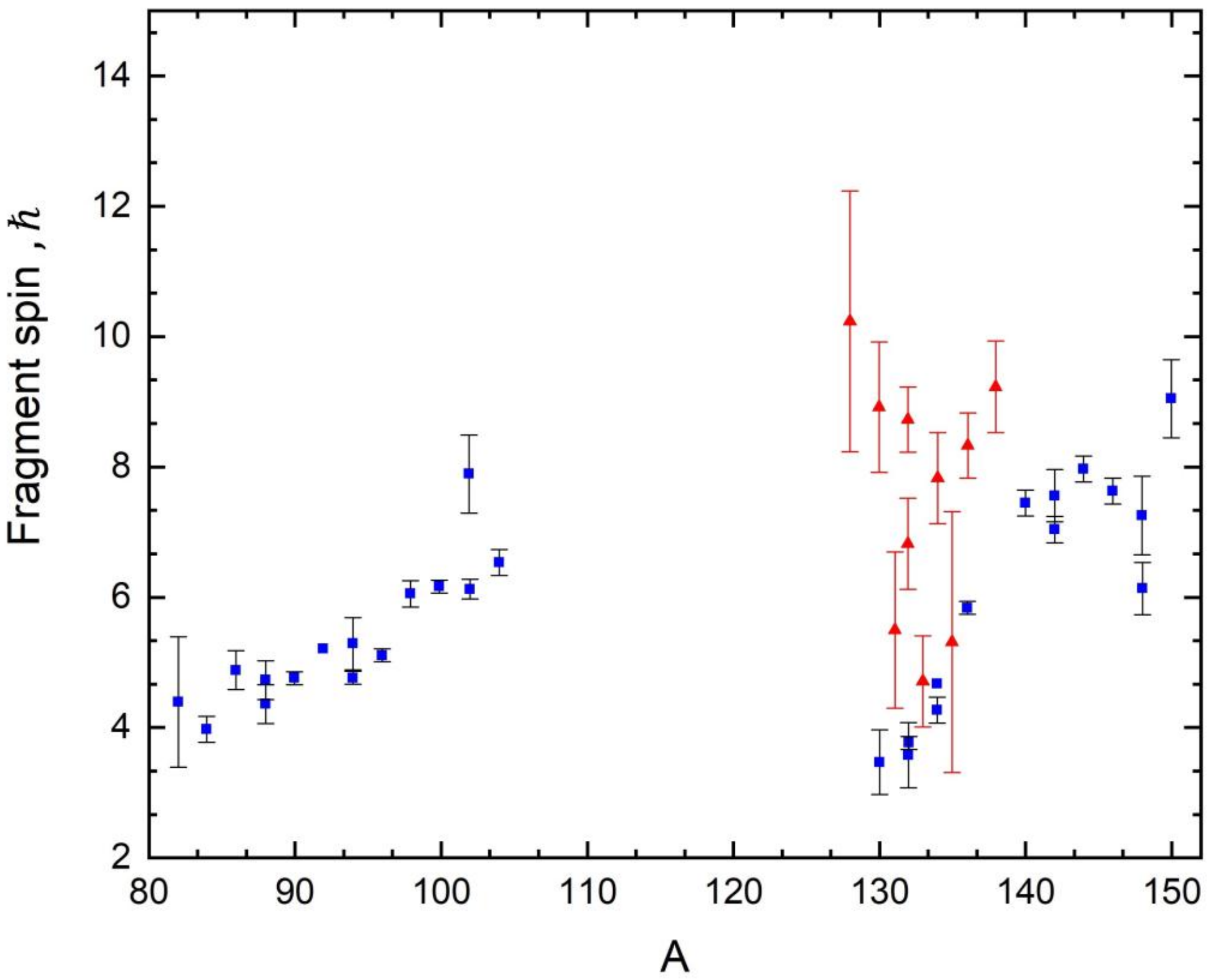

*Fig. 8. Comparison of experimental average spin values as a function of the mass of stimulated fission fragments of* $^{238}\mathrm{U}(n,f)$. *Red triangles indicate results from [32], and blue squares from [1].*

Regarding the isotopes $^{237}\mathrm{Np}$, $^{233}\mathrm{U}$, $^{234}\mathrm{U}$, $^{235}\mathrm{U}$, $^{238}\mathrm{Pu}$, $^{239}\mathrm{Pu}$, $^{240}\mathrm{Pu}$, $^{241}\mathrm{Pu}$, $^{242}\mathrm{Pu}$, modern SD data for these isotopes are not currently available, although similar data are present in earlier studies [31]. Because the results presented in these works cannot be confirmed using modern experiments such as [1], only up-to-date experimental data have been used to validate the models in this study. Thus, the lack of up-to-date and reliable SD data for other isotopes, including $^{237}\mathrm{Np}$, $^{233}\mathrm{U}$, $^{234}\mathrm{U}$, $^{235}\mathrm{U}$, $^{238}\mathrm{Pu}$, $^{239}\mathrm{Pu}$, $^{240}\mathrm{Pu}$, $^{241}\mathrm{Pu}$, $^{242}\mathrm{Pu}$, limited the scope of our study, which focused on the forced threshold fission of nuclei $^{232}\mathrm{Th}$ and $^{238}\mathrm{U}$, for which the most reliable experimental data are available.

## CONCLUSION

This study extends methodological approaches originally developed for spontaneous fission of $^{252}Cf$ [18] to neutron-induced threshold fission of $^{232}Th$ and $^{238}U$ nuclei. A comprehensive investigation of fission fragment moment of inertia behavior has been conducted, enabling the identification of characteristic features inherent to different theoretical models employed for their description. The performed analysis yields several fundamental conclusions bearing significant implications for fission theory development.

The developed indirect methodology, based on comparative analysis of fragment mean spin values, demonstrates the inadequacy of superfluid models employing both oscillator and rectangular potentials under conditions of substantial non-equilibrium deformations. Our analysis unequivocally establishes that achieving consistent agreement between theoretical predictions and experimental data [1] for both nuclei across the complete mass range of light and heavy fragments (A = 80-150) necessitates implementation of hydrodynamic approach for moment of inertia estimation.

These findings are consistent with conclusions from previous comprehensive research on fragment moments of inertia in spontaneous fission [18], which established hydrodynamic model dominance throughout the mass number range. For threshold fission of $^{232}Th$ and $^{238}U$ nuclei, an analogous relationship between equilibrium and non-equilibrium deformation parameters persists across the entire investigated mass region. Consequently, optimal theory-experiment correspondence is achieved through hydrodynamic moment of inertia model implementation, mirroring results obtained for spontaneous fission.

Under conditions of small quadrupole deformations approaching equilibrium values, an adequate description is provided by Cooper pairing and superfluid nucleon-nucleon correlations. The superfluid nuclear model with oscillator potential yields satisfactory results in these regimes. However, upon fragment transition to non-equilibrium deformation domains (reaching anomalously high values approaching unity), where mean free path becomes smaller than nuclear dimensions, the dominant description is provided by the hydrodynamic model with the potential character of collective nucleon motion. These conclusions are confirmed through a detailed analysis of Fig. 5-6 and Tables I-II.

Consequently, the research findings demonstrate the necessity of employing distinct theoretical frameworks contingent upon the nuclear deformation magnitude. The obtained conclusions hold considerable significance for both the broad nuclear physics community and computational software developers, particularly authors of the FREYA code implementing "at hot" moment of inertia approximations for spin distribution calculations. This investigation establishes the superiority of the presented approaches over conventional approximations, thereby providing foundations for substantial refinement of existing computational methodologies.

The singular importance of these results lies in conducting the first comprehensive comparative analysis of nuclear moment of inertia models utilizing diverse theoretical frameworks, yielding novel insights into their formation mechanisms. Further investigations examining model applicability across broader nuclear ranges are required to broaden our understanding of nuclear fission. Such endeavors will not only refine respective domains of model adequacy but also expand the frontiers of contemporary quantum fission theory.

A logical next step would be to apply this methodology to nuclei undergoing thermal neutron fission, which are crucial for reactor energy production — specifically isotopes $^{233}U$, $^{235}U$, $^{239}Pu$, and $^{241}Pu$. However, meticulous analysis has revealed significant discrepancies between late 20th-century compound fission system spin distributions and contemporary experimental data. Early studies of these isotopes [$^{237}Np$, $^{233}U$, $^{234}U$, $^{235}U$, $^{238}Pu$, $^{239}Pu$, $^{240}Pu$, $^{241}Pu$, $^{242}Pu$] exhibit systematic overestimation of spin values, raising serious concerns regarding their reliability. The contradiction between elevated spin values for fission fragments of near-magic actinide nuclei and modern theoretical frameworks [1] necessitated an investigation on transition toward higher energy regimes, specifically threshold fission of $^{232}Th$ and $^{238}U$ induced by 2 MeV neutrons.

Contemporary experimental data [1] for $^{232}$Th and $^{238}$U nuclei demonstrate excellent agreement with current theoretical models. Because these systems represent highly excited compound nuclei, this agreement indirectly validates our methodological approach for less-excited systems. Nevertheless, definitive verification requires additional experimental investigations, potentially utilizing apparatus previously employed by the Wilson group for spontaneous $^{252}$Cf fission and threshold fission studies of $^{232}$Th and $^{238}$U [1].

In summary, this research provides significant contributions to understanding fission fragment internal structure and moment of inertia formation mechanisms, thereby creating new avenues for investigating nuclear reactions and their applications across diverse technological and scientific domains.

## REFERENCES

1. J. Wilson et al., *Nature* **590**, 566 (2021).
2. I. Stetcu et al., *Phys. Rev. Lett.* **127**, 222502 (2021).
3. J. Randrup and R. Vogt, *Phys. Rev. Lett.* **127**, 062502 (2021).
4. A. Bohr and B. Mottelson, *Nuclear Structure*, Benjamin, 1977.
5. S.G. Kadmensky and V.I. Furman, *Alfa-raspad i rodstvennye yadernye reaktsii* (Energoatomizdat, 1985).
6. E.P. Wigner, *Ann. Math.* **62**, 548 (1955); **65**, 203 (1958); **67**, 325 (1958).[6,7,8]
7. D.E. Lyubashevsky, A.A. Pisklyukov, S.V. Klyuchnikov, and P.V. Kostryukov, Estimation of correlation coefficients and spin angular distributions of fission fragments, *Phys. Rev. C* **111**, 054601 (2025). DOI:10.1103/PhysRevC.111.054601.[9]
8. S.G. Kadmensky, V.P. Markushev, V.I. Furman, *Yad. Fiz.* **31**, 382 (1980).[10]
9. S.G. Kadmensky and L.V. Titova, *Phys. Atom. Nucl.* **72**(10), 1738–1744 (2009).[11]

10. D.E. Lyubashevsky, J.D. Shcherbina, and S.G. Kadmensky, Quantum nature of P-even T-odd asymmetries in differential cross sections of fission reactions of unpolarized target nuclei by cold polarized neutrons with escape of prescission and evaporation of light particles, *Phys. Rev. C* **111**, 024609 (2025).[12]
11. A. Gagarski, F. Goennenwein, I. Guseva, et al., *Phys. Rev. C* **93**, 054619 (2016).[13]
12. G.V. Danilyan, *Phys. Atom. Nucl.* **82**, 250 (2019).[14]
13. V. Strutinsky, *Nucl. Phys. A* **95**, 420 (1967).[15]
14. A. Tudora, *Nucl. Phys. A* **916**, 79–101 (2013).[16]
15. A. Tudora, F.-J. Hambsch, and V. Tobosaru, *Phys. Rev. C* **94**, 044601 (2016).[17]
16. D.E. Lyubashevsky, P.V. Kostryukov, A.A. Pisklyukov, J.D. Shcherbina, Evaluation of fission fragment moments of inertia for spontaneous fission of Cf-252, *Chinese Phys. C* **49**(3), 034104 (2025). DOI:10.1088/1674-1137/ad8d4d.[18]
17. C. Hagmann, J. Verbeke, R. Vogt, and J. Randrup, Fission Reaction Event Yield Algorithm, Tech. Rep., Lawrence Livermore National Laboratory (2016).[19]
18. O. Grudzevich, *Problems of Atomic Science and Technology, Series: Nuclear Constants* **39** (2000).[20]
19. R. Walsh and J. Boldeman, *Nucl. Phys. A* **276**, 189 (1977).[21]
20. T. Døssing, S. Åberg, M. Albertsson, B.G. Carlsson, and J. Randrup, *Phys. Rev. C* **109**, 034615 (2024).[22]
21. P. Möller, A.J. Sierk, T. Ichikawa, and H. Sagawa, *At. Data Nucl. Data Tables* **109**, 1 (2016).[23]
22. A. Sitenko and V. Tartakovskii, *Lectures on the Theory of the Nucleus*, Elsevier, 2014.[24]
23. S. Kadmensky, D. Lyubashevsky, D. Stepanov, and A. Pisklyukov, *Phys. Atom. Nucl.* **87**, 359 (2024).[25]

24. A. Tudora, F.-J. Hambsch, and V. Tobosaru, *Eur. Phys. J. A* **54**, 87 (2018).[26]

25. J. Randrup, T. Døssing, and R. Vogt, *Phys. Rev. C* **106**, 014609 (2022).[27]

26. R. Vogt and J. Randrup, *Phys. Rev. C* **103**, 014610 (2021).[28]

27. A.B. Migdal, *Sov. Phys. JETP* **10**, 176 (1960).[29]

28. G.G. Adamian, N.V. Antonenko, R.V. Jolos, Yu.V. Palchikov, T.M. Shneidman, and W. Scheid, *Phys. Atom. Nucl.* **70**(8), 1350–1356 (2007).[30]

29. S. Yoon, H. Seo, Y.-S. Kim, C. Lee, and H.-D. Kim, *J. Radioanal. Nucl. Chem.* **330**, 481–491 (2021).[31]

30. H. Naik, S.P. Dange, and R.J. Singh, *Phys. Rev. C* **71**, 014304 (2005).[32]

References

[1] J. Wilson et al., Nature 590, 566 (2021)

[2] I. Stetcu et al., Phys. Rev. Lett. 127, 222502 (2021) J. Randrup and R. Vogt, Phys. Rev. Lett. 127, 062502 (2021)

[3] A. Bohr and B. Mottelson, Nuclear Structure, Benjamin, 1977

[4] S. G. Kadmensky and V. I. Furman, Alfa-raspad i rodstvennye yadernye reaktsii (Energoatomizdat, 1985)

[5] S. G. Kadmensky and V. I. Furman, Alfa-raspad i rodstvennye yadernye reaktsii (Energoatomizdat, 1985)

[6] E. P. Wigner, Ann. Math. 62, 548 (1955)

[7] E. P. Wigner, Ann. Math. 65, 203 (1958)

[8] E. P. Wigner, Ann. Math. 67, 325 (1958)

[9] S. G. Kadmensky, V. P. Markushev, and V. I. Furman, Yad. Fiz. 31, 382 (1980)

[10] S. G. Kadmensky and L. V. Titova, Phys. Atom. Nucl. 72, 1738 (2009)

[11] D. E. Lyubashevsky, J. D. Shcherbina, and S. G. Kadmensky, Phys. Rev. C 111, 024609 (2025)

[12] A. Gagarski, F. Goennenwein, I. Guseva et al., Phys. Rev. C 93, 054619 (2016)

[13] A. Gagarski, F. Goennenwein, I. Guseva et al., Phys. Rev. C 93, 054619 (2016)

[14] G. V. Danilyan, Phys. Atom. Nucl. 82, 250 (2019)

[15] V. Strutinsky, Nucl. Phys. A 95, 420 (1967)
[16] A. Tudora, Nucl. Phys. A 916, 79 (2013)
[17] A. Tudora, F. -J. Hambsch, and V. Tobosaru, Phys. Rev. C 94, 044601 (2016)
[18] D. E. Lyubashevsky, P. V. Kostryukov, A. A. Pisklyukov et al., Chin. Phys. C 49(3), 034104 (2025)
[19] C. Hagmann, J. Verbeke, R. Vogt et al., Fission Reaction Event Yield Algorithm, Tech. Rep., Lawrence Livermore National Laboratory (2016)
[20] O. Grudzevich, Problems of Atomic Science and Technology, Series: Nuclear Constants 39 (2000)
[21] R. Walsh and J. Boldeman, Nucl. Phys. A 276, 189 (1977)
[22] T. Døssing, S. Åberg, M. Albertsson et al., Phys. Rev. C 109, 034615 (2024)
[23] P. Möller, A. J. Sierk, T. Ichikawa et al., At. Data Nucl. Data Tables 109, 1 (2016)
[24] A. Sitenko and V. Tartakovskii, Lectures on the Theory of the Nucleus (Elsevier, 2014)
[25] S. Kadmensky, D. Lyubashevsky, D. Stepanov et al., Phys. Atom. Nucl. 87, 359 (2024)
[26] J. Randrup, T. Døssing, and R. Vogt, Phys. Rev. C 106, 014609 (2022)
[27] R. Vogt and J. Randrup, Phys. Rev. C 103, 014610 (2021)
[28] A. B. Migdal, Sov. Phys. JETP 10, 176 (1960)
[29] G. G. Adamian, N. V. Antonenko, R. V. Jolos et al., Phys. Atom. Nucl. 70, 1350 (2007)
[30] S. Yoon, H. Seo, Y. -S. Kim et al., J. Radioanal. Nucl. Chem. 330, 481 (2021)
[31] H. Naik, S. P. Dange, and R. J. Singh, Phys. Rev. C 71, 014304 (2005)
[32] A. Tudora, F. -J. Hambsch, and V. Tobosaru, Eur. Phys. J. A 54, 87 (2018)